\documentclass{nature}

\usepackage{graphicx}
\makeatletter
\let\saved@includegraphics\includegraphics
\AtBeginDocument{\let\includegraphics\saved@includegraphics}
\renewenvironment*{figure}{\@float{figure}}{\end@float}
\makeatother
\usepackage[separate-uncertainty=true]{siunitx}
\usepackage{subcaption}
\usepackage{amssymb}
\usepackage{amsmath}
\usepackage{enumitem}
\usepackage{bbold}
\usepackage{color}
\usepackage{lineno}
\usepackage{ulem}
\usepackage{wasysym}
\usepackage{multirow}
\usepackage{pdflscape}
\usepackage{booktabs}
\usepackage{comment}

\title{Truly chiral phonons in $\alpha$-HgS}

\author{Kyosuke~Ishito$^{1\#}$, Huiling~Mao$^{1\#}$, Yusuke~Kousaka$^{2,3}$, Yoshihiko~Togawa$^{2}$, Satoshi~Iwasaki$^{3}$, Tiantian~Zhang$^{1,4}$, Shuichi~Murakami$^{1,4}$, Jun-ichiro~Kishine$^{5,6}$ \& Takuya~Satoh$^{1*}$}

\begin{document}

	\maketitle

	\begin{affiliations}
		\item Department of Physics, Tokyo Institute of Technology, Tokyo 152-8551, Japan
		\item Department of Physics and Electronics, Osaka Prefecture University, Osaka 599-8531, Japan
		\item Research Institute for Interdisciplinary Science, Okayama University, Okayama 700-8530, Japan
		\item Tokodai Institute for Element Strategy (TIES), Tokyo Institute of Technology, Tokyo 152-8551, Japan
		\item Division of Natural and Environmental Sciences, The Open University of Japan, Chiba 261-8586, Japan
		\item Institute for Molecular Science, Okazaki, Aichi 444-8585, Japan \\
		$^\#$These authors contributed equally to this work \\
		$^*$e-mail: satoh@phys.titech.ac.jp
	\end{affiliations}

	\begin{abstract}
		Chirality is a manifestation of the asymmetry inherent in nature.
		It has been defined as the symmetry breaking of the parity of static objects\cite{Kelvin04}, and the definition was extended to dynamic motion such that true and false chiralities were distinguished\cite{Barron09}.
		Recently, rotating, yet not propagating, atomic motions were predicted and observed in two-dimensional materials, and they were referred to as `chiral phonons' \cite{Zhang15,Chen19,Zhu18}.
		A natural development would be the discovery of truly chiral phonons that propagate while rotating in three-dimensional materials.
		Here, we used circularly polarised Raman scattering and first-principles calculations to identify truly chiral phonons in chiral bulk crystals.
		This approach enabled us to determine the chirality of a crystal in a non-contact and non-destructive manner.
		In addition, we demonstrated that the law of the conservation of pseudo-angular momentum holds between circularly polarised photons and chiral phonons.
		These findings are expected to help develop ways for transferring the pseudo-angular momentum from photons to electron spins via the propagating chiral phonons in opto-phononic-spintronic devices.
	\end{abstract}


	Chirality, the breaking of reflection and inversion symmetries\cite{Kelvin04}, appears at all levels of nature and has been studied in many fields of research\cite{Cahn66}.
	Barron extended the concept of chirality to the dynamic regime
	by classifying into \textit{true} and \textit{false} chiralities\cite{Barron09}.
	True chirality is demonstrated by systems that exist in two different enantiomeric states that are interconverted by spatial inversion $( \mathcal{P})$, rather than by time inversion $(\mathcal{T})$ combined with appropriate spatial rotation $(\mathcal{R})$\cite{Barron09}. This concept contrasts with false chirality,
	where the two states are interconverted by $\mathcal{P}$ and by $\mathcal{T}$; for example, the magneto-optical Faraday rotation.

	Chirality was further extended to the dynamics of quasi-particles. Possible band-structure shapes have been classified in terms of pseudo-momentum and pseudo-angular momentum (PAM) \cite{Bozovic84},
	which originate from the phase factors
	acquired by discrete translation and rotation symmetry operations on wave functions, respectively.
	In crystals, they are different from the momentum and angular momentum (AM), which are continuous linear and circular motions of an object in real space, respectively.
	Recently, circularly rotating atomic motion with nonzero PAM and AM at high symmetry points in the Brillouin zone in monolayer hexagonal lattices was theoretically predicted and named `chiral phonons' \cite{Zhang15,Chen19}. This prediction was experimentally verified in the transition metal dichalcogenide WSe$_{2}$ via transient infrared (IR) spectroscopy\cite{Zhu18}.

	Employing Barron's true chirality, one can extend the concept of a chiral phonon to `truly chiral phonons', which have two enantiomeric chiral
	modes that propagate with finite group velocity while circularly rotating in three-dimensional materials wherein PAM and AM have opposite signs. These modes should be interconverted by $\mathcal{P}$, but not by $\mathcal{RT}$.
	The observation of truly chiral phonons would require the use of chiral materials, as has been proposed theoretically\cite{Kishine20}.
	Several Raman experiments designed to identify phonon symmetry in chiral crystals were conducted \cite{Pine69,Zallen70,Pine71,Grimsditch77,Imaino80,Garasevich95,Pinan99}. Frequency-degenerated phonon modes were observed to undergo splitting at the centre of the Brillouin zone ($\Gamma_3$ doublets) in $\alpha$-quartz and trigonal tellurium \cite{Pine69,Pine71}. Moreover, theoretical work on phonon band dispersion has been reported on $
	\alpha$-HgS (cinnabar), where the splitting was attributed to the linear wave-vector shifts caused by chirality \cite{Cardona10}. Nonetheless, previous studies have mostly focused on examining the phonon symmetry without considering PAM and AM.

	Here, we used circularly polarised Raman spectroscopy (see Methods) to observe the phonon modes and their splitting in chiral crystals of $\alpha$-HgS.
	Moreover, we performed first-principles calculations (see Methods) to compute the dispersion, PAM, and AM of phonons in $\alpha $-HgS. A comparison between the experiment and calculation shows that the split phonons were indeed truly chiral phonons.
	We also confirmed that the conservation law of PAM holds in circularly polarised Raman scattering by taking into account the three-fold rotational symmetry.


	Trigonal $\alpha $-HgS crystals contain two enantiomeric cinnabar structures with right- and left-handed helices (R- and L-HgS), which belong to the space groups $P3_{1}21$ (translation by $c/3$ in three-fold rotation) and $P3_{2}21$ (translation by $2c/3$ in three-fold rotation), respectively \cite{Aurivillius50}.
	This crystal has a three-fold helical axis along the $c$-direction (Fig. 1(a)), with large optical activity\cite{Glazer86}.
	Our samples were single crystals of naturally grown $\alpha$-HgS with lateral dimensions of 3 mm. We selected a $c$-axis-oriented as-grown surface of R-HgS, which was confirmed by X-ray analysis.


	The phonon modes at the centre of the Brillouin zone ($\Gamma$ point) of
	$\alpha $-HgS are classified into two $\Gamma_1^{(1,2)}$ singlets (Raman active), three $\Gamma_2^{(1\textendash3)}$ singlets (IR active), and five $\Gamma_3^{(1\textendash5)}$ doublets (Raman and IR active).
	Our measurement led to the observation of the $\Gamma_1^{(1,2)}$ singlets and $\Gamma_3^{(2\textendash5)}$ doublets, which were assigned on the basis of the selection rule\cite{Higgs53} for the four incident/scattered polarisation configurations (RR, LL, RL, and LR), where $I_\text{RR}: I_\text{LL}: I_\text{RL}: I_\text{LR}=1:1:0:0$ and $0:0:1:1$ for the $\Gamma_1$ and $\Gamma_3$ modes, respectively (Extended Data Fig. 1).
	The transverse optical (TO) phonons of the $\Gamma_3$ mode were generated because the incident and scattered light were propagated along the $c$-axis. A weak signal was detected for the longitudinal optical (LO) phonon at $\sim 145$ cm$^{-1}$, presumably due to misalignment of the experimental setup.

	The phonon frequencies obtained by the Raman experiment are compared with the results of the first-principles calculations and with previous results\cite{Zallen70,Cardona10} in Extended Data Table 1.
	The results of the Raman experiments both of our study and of a previous study~\cite{Zallen70} are identical to within 3 cm$^{-1}$. The results of our calculations at the $\Gamma$ point also closely approximate those in Ref.~\cite{Cardona10}, and the small differences are attributed to the difference in the positions of the atoms.
	In our experiments, the $\Gamma_3^{(1)}$ mode was not observed because this mode is nearly degenerate with the $\Gamma_1^{(1)}$ mode, which has a much higher intensity.
	In Extended Data Table 1, the frequencies of the $\Gamma_3^{(2\textendash5)}$ modes obtained in our Raman experiment are averaged for the RL and LR configurations, which were split as shown in the next paragraph.

	The Stokes and anti-Stokes Raman spectra (Raman optical activity) of each $\Gamma_3^{(2\textendash5)}$ mode are shown in Figs. 2(a--h). All four $\Gamma_3$ modes were observed with the opposite circularly polarised light between the incident and scattered light (RL and LR). On the anti-Stokes spectra [see Figs. 2(a--d)], the absolute values of the Raman shift of the $\Gamma_3^{(2)}$ and $\Gamma_3^{(4)}$ modes in the RL configuration are lower than those in the LR configuration in Figs. 2(a) and (c). In Figs. 2(b) and (d), the absolute values of the Raman shift of the $\Gamma_3^{(3)}$ and $\Gamma_3^{(5)}$ modes in the RL configuration are higher than those in the LR configuration. On the Stokes spectra [see Figs. 2(e--h)], the splittings are mirror images of those on the anti-Stokes spectra [see Figs. 2(a--d)]. The $\Gamma_1^{(1,2)}$ modes do not split. 

	The properties of phonons are discussed here on the basis of our first-principles calculations. Figure 3(a) displays the phonon dispersion curve calculated along the path $\Delta$ from $\Gamma$ to A point, which is parallel to the $c$-axis [see Fig. 1(b)]. Linear splittings of $\Gamma_3^{(1-5)}$ modes appear around the $\Gamma $ point, which is consistent with previous calculations\cite{Cardona10,Nusimovici73b}.
	From the conservation law of pseudo-momentum, the wavenumber $k$ of phonons observed by backscattering Raman spectroscopy is expressed as $k=4\pi n/\lambda$, where $\lambda$ is the wavelength of the incident or scattered light, and $n=2.77$ is the absolute value of the complex refractive index of $\alpha $-HgS\cite{Bond67}. The value of $k$ was calculated to be $\sim 1/80$ of the wavenumber $\pi /c$ at the A point. We denoted this specific point on the path $\Delta$ as `N' in Fig. 3(a).
	The experimental splitting values of the $\Gamma_3$ modes at this point (Stokes scattering) are almost consistent with the calculations, as indicated in Extended Data Table 2.
	Therefore, we attribute the splits observed in Fig. 2 to the linear $k$ shifts at the N point.

	To understand the selection rule of the incident/scattered polarisation configurations on the Raman spectra in Fig. 2, we calculated the PAM of phonons in R-HgS (see Methods). As shown in Fig. 3(a), the $\Gamma_3$ doublets have spin PAM $m_{\text{PAM}}^{\text{s}}=\pm 1$, whereas the $\Gamma_1$ and $\Gamma_2$
	singlets have $m_{\text{PAM}}^{\text{s}}=0$. The phonon bands with  $m_{\text{PAM}}^{\text{s}}=-1$ and $m_{\text{PAM}}^{\text{s}}=0$ intersect at the A point.
	These are the common features of the $3_1$ (right-handed) helix\cite{Bozovic84,Pine69,Nussbaum61}.
	From Figs. 2(e--h) and 3(a), $\Gamma_3^{(2\textendash5)}$ phonons with  $m_{\text{PAM}}^{\text{s}}=+1$ and $-1$ were observed in the LR and RL configurations, respectively (Extended Data Table 2).
	$\Gamma_1$ phonons with $m_{\text{PAM}}^{\text{s}}=0$ were observed in the RR and LL configurations.
	Right- and left-handed circularly polarised light possess PAM of $\sigma=+1$ and $-1$, respectively\cite{Yariv07}.
	Therefore, we confirmed a conservation law to exist between the spin PAM of phonons and the PAM of the incident/scattered photons in the Stokes spectra as
	\begin{equation}
	\sigma _{\text{s}}-\sigma _{\text{i}}=-m_{\text{PAM}}^{\text{s}}+3p,  \label{Umklapp}
	\end{equation}
	where $\sigma _{\text{s}}$ and $\sigma _{\text{i}}$ represent the PAM of the scattered and incident photons, respectively. In addition, $p=0$ and $\pm1$ for the $\Gamma_1$ and $\Gamma_3$ modes, respectively.
	The factor $3p$ on the right-hand side of equation (\ref{Umklapp}) can be understood by considering the
	three-fold rotational symmetry of $\alpha$-HgS and the Umklapp process in Raman scattering\cite{Bozovic84,Bloembergen80,Tatsumi18,HChen21,Zhang21}.

	In crystals with discrete rotational symmetry, a clear distinction should be made between the PAM and AM of the phonons.
	The AM of phonons arises from the circular vibration of atoms in the real space, as shown in Figs. 1(d--f) [See also Supplementary Videos 1 and 2].
	The AM of phonons along the $c$-axis is displayed in Fig. 3(b) along with $\Gamma$ to A point (see Methods).
	In the vicinity of the $\Gamma$ point, the $\Gamma_3$ modes have clockwise and  counter-clockwise rotations, corresponding to the positive and negative AM of phonons, respectively. This is a clear manifestation of the existence of chiral phonons.
	The $\Gamma_1$ modes have linear vibrations, and their phonons have zero AM. The signs of AM at the N point are listed in Extended Data Table 2, where the spin PAM and AM of phonons do not correspond with each other. This indicates that, contrary to PAM, the AM of phonons is not conserved during the Raman process in a crystal with discrete rotational symmetry.

	We also calculated the PAM and AM of phonons in L-HgS (Extended Data Fig. 2). The signs of spin PAM and AM are reversed with respect to Fig. 3, that is, $m_{\text{PAM}}^{\text{s}}(\text{RH}, k, j)=-m_{\text{PAM}}^{\text{s}}(\text{LH}, k, j)$ 
	and $m_{\text{AM}}(\text{RH}, k, j)=-m_{\text{AM}}(\text{LH}, k, j)$
	, where RH and LH denote crystals with right- and left-handed helices, respectively.
	This means that the splitting of the $\Gamma_3$ modes arises from chirality.
	From the first-principles calculation, the phonon eigenvector at the N point
	$\textbf{u}(\text{RH}, k, j)$ and $\textbf{u}(\text{LH}, k, j)$ are converted by $\mathcal{P}$, $\mathcal{T}$, and $\mathcal{C}_2$ as
	$\mathcal{P}\textbf{u}(\text{RH}, k, j)=\textbf{u}(\text{LH}, -k, j)$,
	$\mathcal{C}_2\mathcal{T}\textbf{u}(\text{RH}, k, j)=\mathcal{C}_2\textbf{u}(\text{RH}, -k, j)=\textbf{u}(\text{RH}, k, j)$, where $\mathcal{C}_2$ is a two-fold rotation around the $\textbf{a}$, $\textbf{b}$, or $\textbf{a}+\textbf{b}$ direction.
	These relations satisfy the definition of true chirality with two different enantiomeric phonon modes:
	$\textbf{u}(\text{RH}, k, j)$ and $\textbf{u}(\text{LH}, -k, j)$.

	We also recorded the Raman spectra of an $\alpha$-HgS crystal for which the handedness (chirality) of the measured region was unknown because the size of the chiral domain was smaller than the experimental resolution ($\sim$ 1 mm) of our X-ray diffractometer. Contrary to the X-ray analysis, our Raman experiment with a spatial resolution of a few $\mu$m succeeded in observing the splitting of the $\Gamma_3$ modes (Extended Data Fig. 3), and the directions of the splittings were opposite to the results in Fig. 2. Considering the data in Extended Data Table 3, the chirality of the sample at the point of measurement was found to be left-handed.

	The space group analysis enabled us to additionally assign an irreducible representation of the phonon dispersion curve in R-HgS. On path $\Delta$ from $\Gamma$ to A, three irreducible representations are known to exist: $\Delta_1$, $\Delta_2$, and $\Delta_3$\cite{Nusimovici73b,Teuchert74}.
	The labels of the irreducible representations are based on the
convention in Ref. \cite{Teuchert74}.
	The $\Gamma_1$ and $\Gamma_2$ modes at the $\Gamma$ point change to the $\Delta_1$ mode, whereas the $\Gamma_3$ modes split into $\Delta_2$ and $\Delta_3$ modes in accordance with the compatibility relations. In Supplementary Note 1, we present a derivation of the Raman tensors that correspond to the $\Delta_2$ and $\Delta_3$ modes. It follows that the Raman intensities of the $\Delta_2$ and $\Delta_3$ modes are
	$I_\text{RR}: I_\text{LL}: I_\text{RL}: I_\text{LR}=0:0:1:0$ and $0:0:0:1$, respectively. This is consistent with the results of the Raman experiments.
	In other words, spin PAM $m_{\text{PAM}}^{\text{s}}(\text{RH})=0, +1, -1$ correspond to $\Delta_1, \Delta_2, \Delta_3$, respectively, because the phase change by three-fold rotation is the index of each irreducible representation.
	For L-HgS, $m_{\text{PAM}}^{\text{s}}(\text{LH})=0, +1, -1$ correspond to $\Delta_1, \Delta_3, \Delta_2$, respectively\cite{Teuchert74}.

	Finally, we discuss the propagation of the chiral phonons. From Fig. 3, the group velocities of the $\Gamma_3^{(1-5)}$ modes at the N point are calculated as
	$\sim \pm 0.4\text{--}2\times10^3~\text{m}/\text{s}$, which is comparable to the sound velocity of acoustic phonons.
	Note that the nonreciprocal propagation of chiral phonons can be controlled by the PAM of the photon.
	This implies the possibility of transferring the PAM from photons to electron spins via the propagating chiral phonons in opto-phononic-spintronic devices.
	For example, taking advantage of the long coherence of long-wavelength phonons, transferring the PAM from phonon to electron spins may be realised on the macroscopic scale.

	Weyl phonons can exist widely in chiral crystals, as has been predicted by the first-principles calculations and verified by inelastic X-ray scattering\cite{Zhang18,Li21}. We note that the circularly polarised Raman spectroscopy presented in this paper may open the possibility to detect phonons carrying a nonzero Chern number by measuring the phonon PAM, instead of detecting the eigenvalue and eigenvectors of the topological bands.

	We observed chiral phonons in a three-dimensional chiral system using circularly polarised Raman spectroscopy and first-principles calculations.
	The chiral phonons were labelled with spin PAM of $\{+1, -1\}$ corresponding to $\{\Delta_2, \Delta_3\}$ and $\{\Delta_3, \Delta_2\}$ for  R- and L-HgS, respectively, with opposite group velocities of $\sim$ 1~km$/$s. The parity and time-reversal symmetries of the phonons satisfy the definition of truly chiral phonons, which propagate while rotating along the $c$-axis. This is distinct from the chiral phonons observed in two-dimensional hexagonal systems.
	Our work also provides an optical method to identify the handedness of chiral materials using PAM and we demonstrated that the spatial imaging of chiral domains can be achieved in a non-contact and non-destructive manner.
	
	\section*{Acknowledgments}
	We would like to thank M. Kichise, A. Kawano, K. Matsumoto, A. Koreeda, Y. Fujii, E. Oishi, and H. M. Yamamoto for their valuable discussions and technical support. T.S. was financially supported by the Japan Society for the Promotion of Science (JSPS) KAKENHI (Grant No. JP19H01828, No. JP19H05618, No. JP19K21854, No. JP21H01032, No. JP22H01154), and the Frontier Photonic Sciences Project of the National Institutes of Natural Sciences (NINS) (Grant Nos. 01212002 and 01213004).

    \section*{Author contributions}
	T.S. and J.K. conceived the study. K.I. and H.M. performed the Raman experiments.
	Y.K., Y.T., and S.I. conducted the X-ray analyses. K.I., H.M., T.Z., and S.M. performed the first-principles calculations.
	K.I., H.M., J.K., and T.S. wrote the manuscript. All authors discussed the results and commented on the manuscript.
	
	\section*{Competing interests}
	The authors declare no competing interests.

	\newpage
	\begin{table}
		{\textbf{Extended Data Table 1 $\mid$ phonon frequencies in $\alpha$-HgS.} The experimental values in cm$^{-1}$ were obtained at room temperature. The calculated values in cm$^{-1}$ are at the $\Gamma$ point.}

		\begin{center}
			\begin{tabular}{ccccc}
				Symmetry&{Our experiments}&{Experiments \cite{Zallen70}}&{Our calculations}&{Calculations \cite{Cardona10}}\\
				\hline
				$\Gamma_{\text{2},\text{LO}}^{(1)}$&Inactive&&40.2&44.1\\
				$\Gamma_{\text{1}}^{(1)}$&42&45&41.0&39.1\\
				$\Gamma_{\text{3},\text{TO}}^{(1)}$&&&43.5&42.0\\
				$\Gamma_{\text{3},\text{TO}}^{(2)}$&85&88&84.4&83.1\\
				$\Gamma_{\text{3},\text{TO}}^{(3)}$&103&106&114.4&121.4\\
				$\Gamma_{\text{2},\text{LO}}^{(2)}$&Inactive&&149.4&159.6\\
				$\Gamma_{\text{1}}^{(2)}$&254&256&237.5&232.1\\
				$\Gamma_{\text{3},\text{TO}}^{(4)}$&282&283&263.0&259.3\\
				$\Gamma_{\text{3},\text{TO}}^{(5)}$&344&345&320.2&319.2\\
				$\Gamma_{\text{2},\text{LO}}^{(3)}$&Inactive&&339.4&337.7\\
			\end{tabular}
		\end{center}
		\label{phononfrequency}
	\end{table}

	\begin{landscape}
		\newcommand{\tabincell}[2]{\begin{tabular}{#1}#2\end{tabular}}
		\begin{table}
			{\textbf{Extended Data Table 2 $\mid$ Chiral properties of phonons in R-HgS along the right-handed helical axis.} Owing to the roughness of the sample surface, multiple splitting measurements were conducted, with the deviation indicated by the error bars.}

			\begin{center}
				\begin{tabular}{c|ccc|ccccc} \toprule
					\multicolumn{1}{c|}{}&\multicolumn{3}{c|}{Experiment}&\multicolumn{4}{c}{Calculation}\\
					\hline
					\tabincell{c}{Symmetry\\at $\Gamma$ point}&\tabincell{c}{Config.}&\tabincell{c}{Frequency\\$[$cm$^{-1}$$]$}&\tabincell{c}{Splitting\\$[$cm$^{-1}$$]$}&\tabincell{c}{Symmetry\\at N point}&\tabincell{c}{Frequency\\$[$cm$^{-1}$$]$}&\tabincell{c}{Splitting\\$[$cm$^{-1}$$]$}&\tabincell{c}{$m_{\text{PAM}}^{\text{s}}$}&\tabincell{c}{$m_{\text{AM}}$}\\
					\hline
					\multirow{2}*{$\Gamma_3^{(2)}$}&RL&84.9&\multirow{2}*{$0.3\pm 0.01$}&$\Delta_3^{(2)}$&84.2&\multirow{2}*{$0.4$}&$-1$&$-$\\
					&LR&85.3&&$\Delta_2^{(2)}$&84.6&&$+1$&$+$\\
					\hline
					\multirow{2}*{$\Gamma_3^{(3)}$}&LR&103.2&\multirow{2}*{$0.3\pm 0.16$}&$\Delta_2^{(3)}$&114.0&\multirow{2}*{$0.9$}&$+1$&$-$\\
					&RL&103.5&&$\Delta_3^{(3)}$&114.9&&$-1$&$+$\\
					\hline


					\multirow{2}*{$\Gamma_3^{(4)}$}&RL&281.9&\multirow{2}*{$0.4\pm 0.08$}&$\Delta_3^{(4)}$&262.7&\multirow{2}*{$0.7$}&$-1$&$+$\\
					&LR&282.3&&$\Delta_2^{(4)}$&263.4&&$+1$&$-$\\
					\hline
					\multirow{2}*{$\Gamma_3^{(5)}$}&LR&343.5&\multirow{2}*{$0.3\pm 0.05$}&$\Delta_2^{(5)}$&320.0&\multirow{2}*{$0.4$}&$+1$&$+$\\
					&RL&343.7&&$\Delta_3^{(5)}$&320.4&&$-1$&$-$\\
				\end{tabular}
			\end{center}
			\label{R-chirality}
		\end{table}
	\end{landscape}

	\newpage

	\begin{landscape}
		\newcommand{\tabincell}[2]{\begin{tabular}{#1}#2\end{tabular}}
		\begin{table}
			{\textbf{Extended Data Table 3 $\mid$ Chiral properties of phonons in $\alpha$-HgS with unknown chirality.} The error bars take into account that multiple splitting measurements were conducted in the experiment, owing to the roughness of the sample surface. Calculations were performed for L-HgS.}

			\begin{center}
				\begin{tabular}{c|ccc|ccccc} \toprule
					\multicolumn{1}{c|}{}&\multicolumn{3}{c|}{Experiment}&\multicolumn{4}{c}{Calculation}\\
					\hline
					\tabincell{c}{Symmetry\\at $\Gamma$ point}&\tabincell{c}{Config.}&\tabincell{c}{Frequency\\$[$cm$^{-1}$$]$}&\tabincell{c}{Splitting\\$[$cm$^{-1}$$]$}&\tabincell{c}{Symmetry\\at N point}&\tabincell{c}{Frequency\\$[$cm$^{-1}$$]$}&\tabincell{c}{Splitting\\$[$cm$^{-1}$$]$}&\tabincell{c}{$m_{\text{PAM}}^{\text{s}}$}&\tabincell{c}{$m_{\text{AM}}$}\\
					\hline
					\multirow{2}*{$\Gamma_3^{(2)}$}&LR&84.6&\multirow{2}*{$0.4\pm 0.04$}&$\Delta_3^{(2)}$&84.2&\multirow{2}*{$0.4$}&$+1$&$+$\\
					&RL&85.0&&$\Delta_2^{(2)}$&84.6&&$-1$&$-$\\
					\hline
					\multirow{2}*{$\Gamma_3^{(3)}$}&RL&102.9&\multirow{2}*{$0.4\pm 0.04$}&$\Delta_2^{(3)}$&114.0&\multirow{2}*{$0.9$}&$-1$&$+$\\
					&LR&103.3&&$\Delta_3^{(3)}$&114.9&&$+1$&$-$\\
					\hline


					\multirow{2}*{$\Gamma_3^{(4)}$}&LR&281.9&\multirow{2}*{$0.5\pm 0.08$}&$\Delta_3^{(4)}$&262.7&\multirow{2}*{$0.7$}&$+1$&$-$\\
					&RL&282.4&&$\Delta_2^{(4)}$&263.4&&$-1$&$+$\\
					\hline
					\multirow{2}*{$\Gamma_3^{(5)}$}&RL&343.1&\multirow{2}*{$0.4\pm 0.03$}&$\Delta_2^{(5)}$&320.0&\multirow{2}*{$0.4$}&$-1$&$-$\\
					&LR&343.5&&$\Delta_3^{(5)}$&320.4&&$+1$&$+$\\
				\end{tabular}
			\end{center}
			\label{L-chirality}
		\end{table}
	\end{landscape}
	\newpage
	\begin{figure}[htb]
		\begin{center}
			\includegraphics[width=16cm]{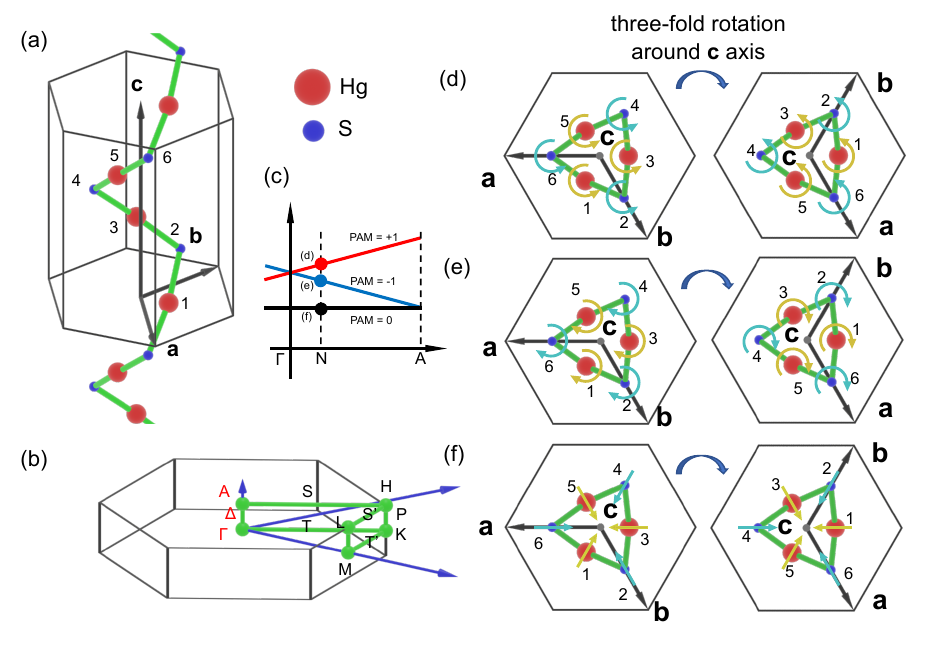}
		\end{center}
		\caption{\textbf{Chiral properties of $\protect\alpha$-HgS.} (a)
			Right-handed helical chain along the $c$-axis of R-HgS. Each Hg atom (red
			sphere) is bonded with two S atoms (blue spheres). Chemical bonds are
			denoted as green lines. (b) Brillouin zone of $\protect\alpha$-HgS. (c)
			Schematic diagram of phonon dispersion from $\Gamma$ to A point. The red,
			blue, and black filled circles at the N point correspond to the phonon modes of panels
			(d), (e), and (f), respectively. Two-dimensional projections of schematic
			atomic motions in R-HgS at point N for (d) a $\Delta_2$ mode with $m_{\text{PAM}}^{\text{s}} = +1$, $m_{\text{AM}} > 0$, (e) a $\Delta_3$ mode with $m_{\text{PAM}}^{\text{s}} = -1$, $m_{\text{AM}} < 0$, and (f) a $\Delta_1$
			mode with $m_{\text{PAM}}^{\text{s}} = 0$ and $m_{\text{AM}} \simeq 0$.
			Counter clockwise, clockwise circular, and linear motions indicate that $m_{\text{AM}}$ is positive, negative, and zero, respectively. The three-fold rotation
			symmetry operation generates phase factors of atomic motions, namely, PAM $m_{\text{PAM}}^{\text{s}}$. The phase difference at the position of each atom after
			the operation is shown. }
		\label{Figure 1}
	\end{figure}

	\newpage
	\begin{figure}[htb]
		\begin{center}
			\includegraphics[width=16cm]{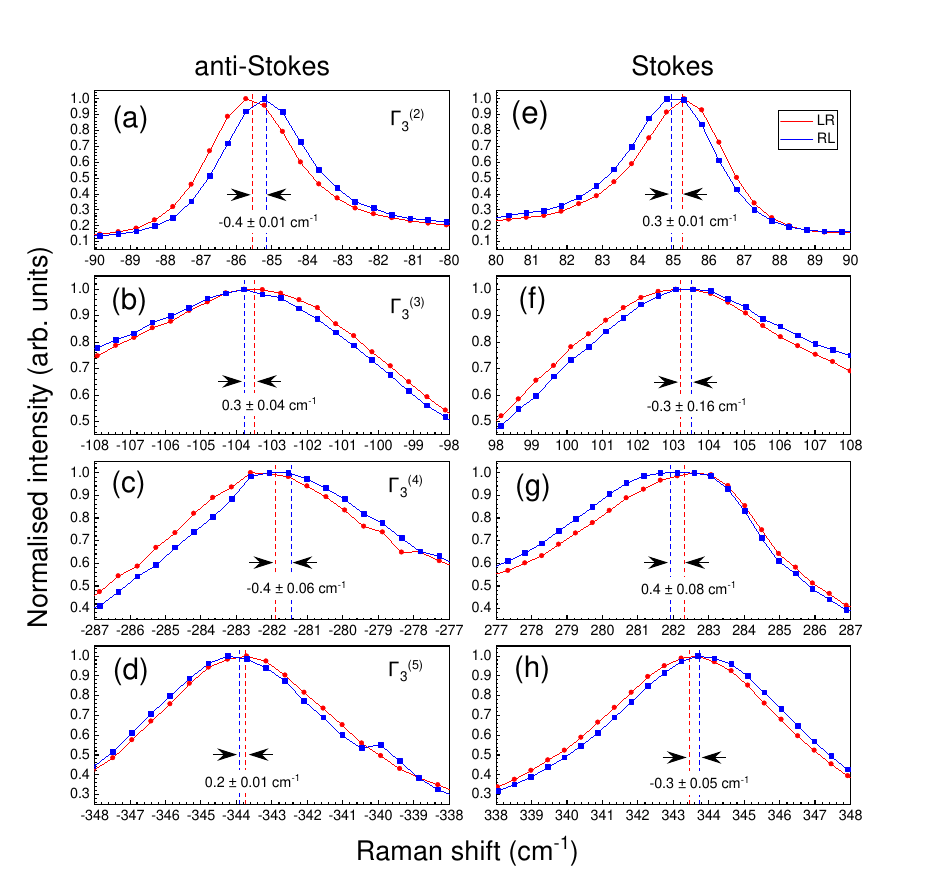}
			\caption{\textbf{Raman spectra of R-HgS.}
				(a--d) Anti-Stokes and (e--h) Stokes spectra of the $\Gamma_3$ modes. The blue and red lines denote the $\Gamma_3$ modes that were experimentally observed with RL- and LR-polarised configurations, respectively. The values of the $\Gamma_3$ doublet splittings are shown in each figure. The error bars take into account that multiple splitting measurements were conducted in the experiment, owing to the roughness of the sample surface.}
		\end{center}
		\label{Figure 2}
	\end{figure}

	\newpage
	\begin{figure}[htb]
		\begin{center}
			\includegraphics[width=16cm]{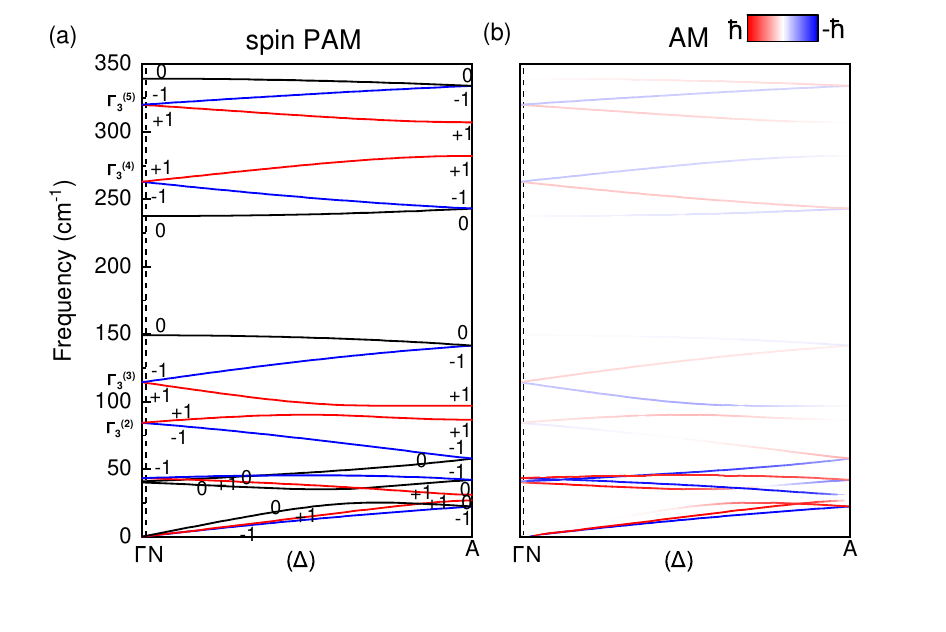}
			\caption{\textbf{Phonon dispersion curve of R-HgS along the right-handed helical axis.} (a) Black, red, and blue curves correspond to spin PAM $m_{\text{PAM}}^{\text{s}}=0, +1, -1$, respectively. The wavenumbers at the N and A points are in the ratio of $1:80$. (b) Phonon AM, denoted by the colour gradient. The red and blue curves correspond to the positive and negative AM, respectively.}
		\end{center}
		\label{Figure 3}
	\end{figure}

	\newpage
	\begin{figure}[htb]
		\begin{center}
			\includegraphics[width=16cm]{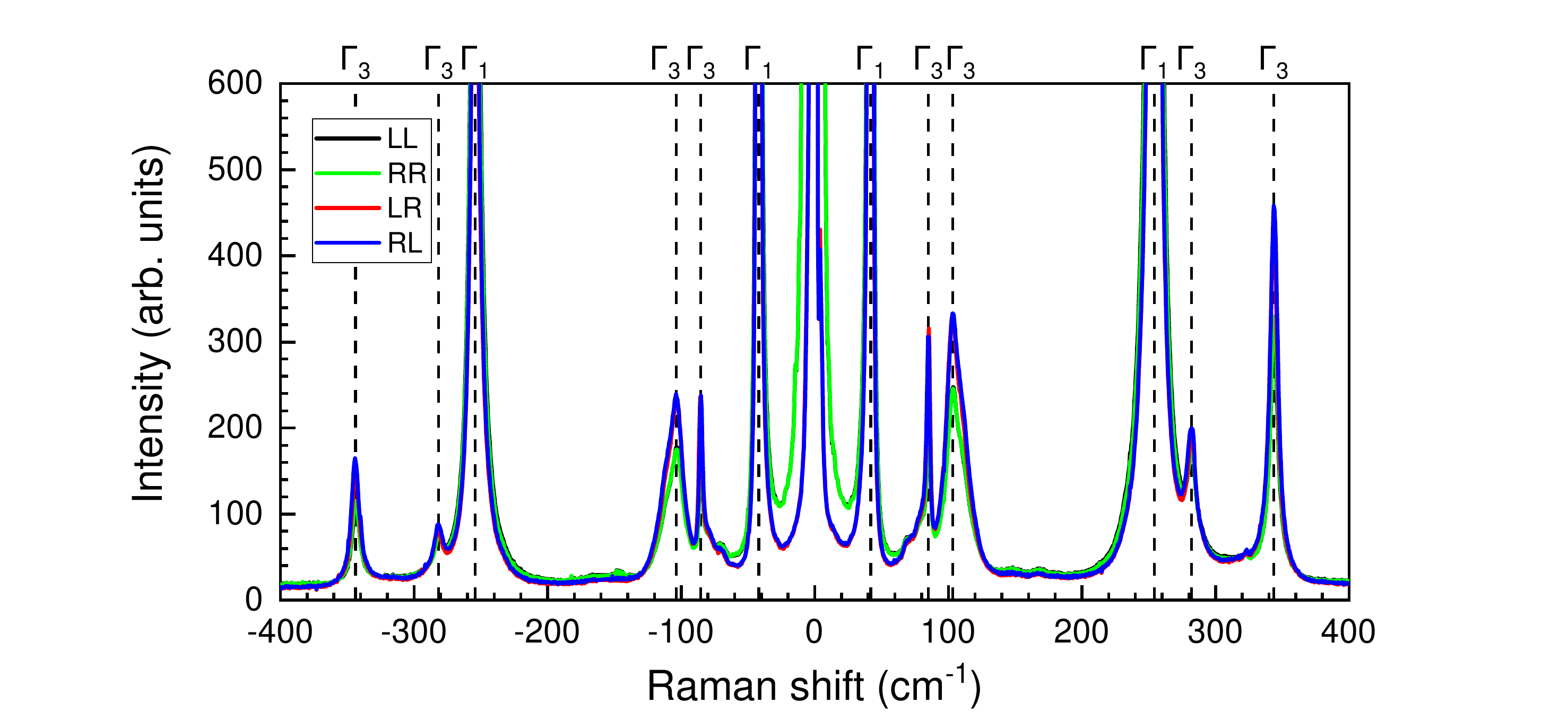}
		\end{center}
		{\textbf{Extended Data Fig. 1 $\mid$ Full Raman spectra of R-HgS.} Full Raman spectra were recorded at room temperature using various polarised configurations. The black, green, red, and blue spectra represent the LL-, RR-, LR-, and RL-polarised configurations, respectively. R and L represent the right- and left-handed helicities of circularly polarised light, respectively.}
	\end{figure}

	\newpage
	\begin{figure}[htb]
		\begin{center}
			\includegraphics[width=16cm]{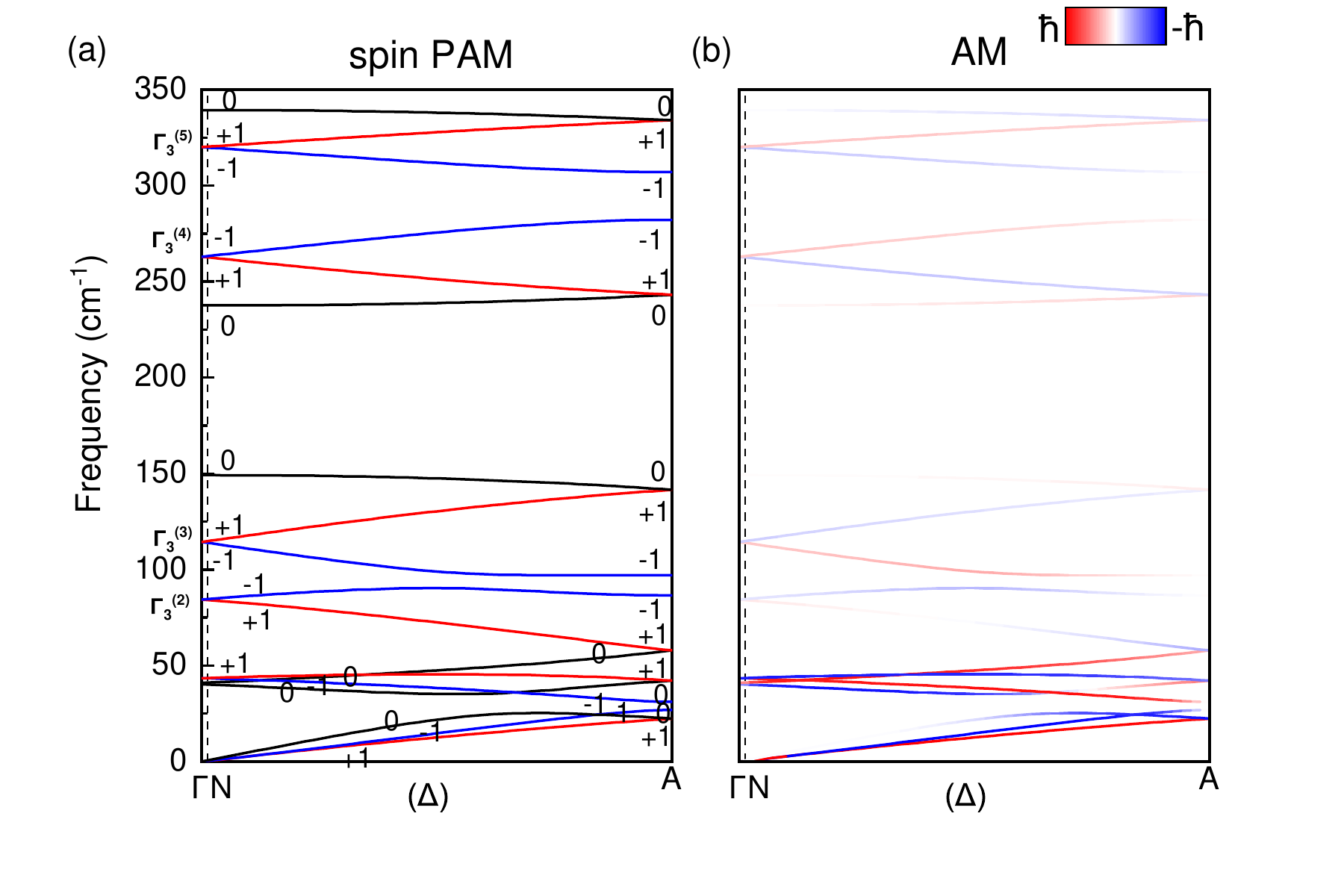}
		\end{center}
		{\textbf{Extended Data Fig. 2 $\mid$ Phonon dispersion curve of L-HgS along the left-handed helical axis.} (a) The black, red, and blue curves correspond to spin PAM $m_{\text{PAM}}^{\text{s}}=0, +1, -1$, respectively. (b) Phonon AM, denoted by the colour gradient. The red and blue curves correspond to the positive and negative AM, respectively.}
	\end{figure}

	\newpage
	\begin{figure}[htb]
		\begin{center}
			\includegraphics[width=16cm]{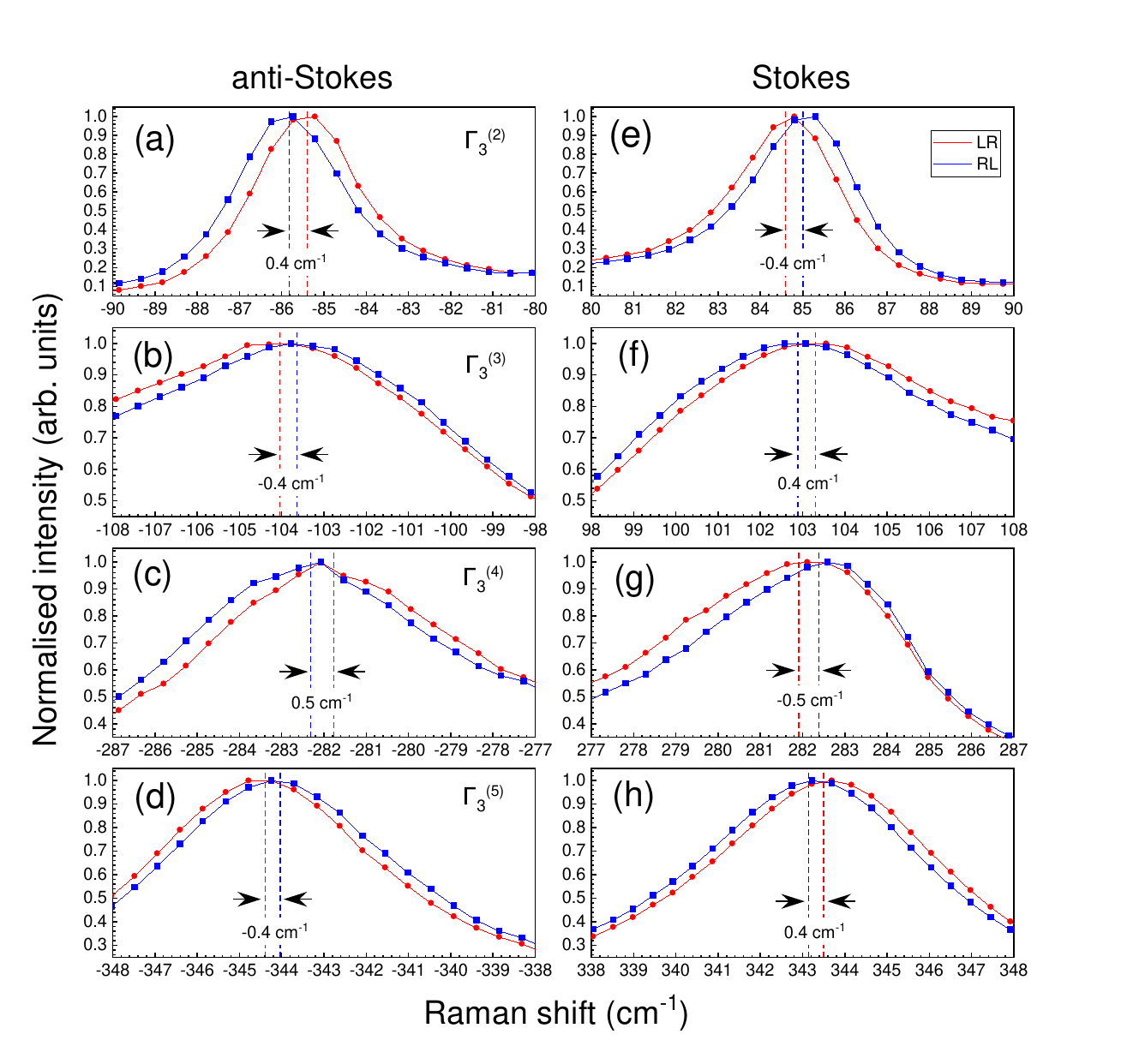}
		\end{center}
		{\textbf{Extended Data Fig. 3 $\mid$ Raman spectra of $\alpha$-HgS with unknown chirality.}
			(a--d) Anti-Stokes and (e--h) Stokes spectra of the $\Gamma_3$ modes. The blue and red lines denote the $\Gamma_3$ modes that were experimentally observed with RL- and LR-polarised configurations, respectively. R and L represent the right- and left-handed helicities of circularly polarised light, respectively. The values of the $\Gamma_3$-doublet splittings are shown in each figure.}
	\end{figure}


	\section*{Methods}
	\textbf{Raman scattering.}
	We used a custom-built Raman spectroscopy system\cite{Hsu20} equipped with a monochromator, charge-coupled device (CCD) camera, and optical
	elements. Details of the setup can be found in Supplementary Fig. 2. The excitation light was generated by a diode laser with a wavelength of $\lambda =785$ nm. The incident and scattered light were propagated along the $c$-axis of the crystal.
	All the measurements were performed at room temperature.

	\noindent \textbf{First-principles calculation.}
	First-principles calculations were performed using the ABINIT package\cite{Gonze09,Romero20}, which implements 
	density functional perturbation theory (DFPT)\cite{Hamann05}. An anaddb was used to obtain the phonon dispersion curve\cite{Gonze97,Gonze97a}. For the calculation of phonon dispersion, the exchange-correlation energy of electrons is described as
	the generalized gradient approximation (GGA) with the norm-conserving
	pseudo-potentials \cite{Setten18,Hamann13}, and spin-orbit coupling was not considered. We set the kinetic energy cut-off to 60 Ha,
	which was necessary to obtain a well-converged result, and the grids of the wave
	vector \textit{k } $6\times 6\times 4$ to describe the phonon dispersion
	curve. In addition, we increased the precision of the \textit{k} grids to $8\times 8\times 8$ to obtain the converged phonon frequency at the $\Gamma$ point. The following lattice parameters of $\alpha$-HgS were used :
	the volume of the cell is $V_{0}=160.02$ \AA $^{3}$, the length of the cell is $a=4.11
	$ \AA $,$ $b=4.11$ \AA $,$ and $c=9.40$ \AA , and the angle of the cell is $\alpha =90^{\circ },$ $\beta =90^{\circ },$ and $\gamma =120^{\circ }$\cite{Jain13,Rodic96}. 
	R-HgS is composed of three Hg $(0.699, 0.699, 0.000)$, $(0.000, 0.301, 2/3)$, $(0.301, 0.000, 1/3)$ 
	and three S $(0.509, 0.509, 1/2)$, $(0.000, 0.491, 1/6)$, $(0.491, 0.000, 5/6)$,
	and L-HgS is composed of three Hg $(0.699, 0.699, 0.000)$, $(0.301, 0.000, 2/3)$, $(0.000, 0.301, 1/3)$ 
	and three S $(0.509, 0.509, 1/2)$, $(0.491, 0.000, 1/6)$, $(0.000, 0.491, 5/6)$, of which the bases are the $a, b, c$ components.
	The essential part of the ABINIT results was confirmed using the VASP package.

	\noindent \textbf{Calculation of PAM.}
	In a material with three-fold rotational symmetry, PAM is defined as follows\cite{Bozovic84,Zhang15}
	\begin{equation}
	\left\{C_3~|~0\right\} \textbf{u}(\textbf{k})=\exp \left[-i\frac{2\pi}{3}
	m_{\text{PAM}}(\textbf{k})\right]\textbf{u}(\textbf{k}),
	\label{rotation function}
	\end{equation}
	where $m_{\text{PAM}}(\textbf{k})$ is the PAM of phonons, $\textbf{u}(\textbf{k})$ is the displacement vector, and $\left\{C_3~|~0\right\}$ is the three-fold rotation around the $c$-axis. However, $\alpha$-HgS has a three-fold helical rotational symmetry instead of a three-fold rotational symmetry. Therefore, we defined the PAM of phonons in R- and L-HgS as follows\cite{Zhang21}:
	\begin{equation}
	\left\{C_3~|~\mathbf{c}/3\right\} \textbf{u}(\text{RH}, \textbf{k})=\exp \left[-i\frac{2\pi}{3} 
	m_{\text{PAM}}(\text{RH}, \textbf{k}) \right]\textbf{u}(\text{RH}, \textbf{k}),
	\label{screw function RH}
	\end{equation}
	\begin{equation}
	\left\{C_3~|~2\mathbf{c}/3\right\} \textbf{u}(\text{LH}, \textbf{k})=\exp \left[-i\frac{2\pi}{3}
	m_{\text{PAM}}(\text{LH}, \textbf{k}) \right]\textbf{u}(\text{LH}, \textbf{k}),
	\label{screw function LH}
	\end{equation}
	where $\left\{C_3~|~\mathbf{c}/3\right\}$ and $\left\{C_3~|~2\mathbf{c}/3\right\}$ are combinations of the three-fold rotation around the $c$-axis and the $c/3$ and $2c/3$ translations along the $c$-axis, respectively.

	The displacement vector of phonons can be expressed as follows
	\begin{equation}
	\textbf{u}_{\kappa}(\textbf{k}, l, j) =m_{\kappa}^{-\frac{1}{2}}\boldsymbol \epsilon_{\kappa} (\textbf{k}, j) \exp \left[i\left\{\textbf{k} \cdot \textbf{R}_{l}-\omega(\textbf{k}, j) t\right\}\right],  \label{wave function}
	\end{equation}
	where $j$ is the number of phonon modes, $m_{\kappa}$ is the mass of the $\kappa$-th atom in the $l$-th unit cell, $\boldsymbol \epsilon_{\kappa} (\textbf{k}, j)$ is the eigenvector of the dynamical matrix, and $\textbf{R}_{l}$ is the position of the $l$-th unit cell.
	The use of equation (\ref{wave function}) indicates that the phase of the eigenvector consists of two parts: $\boldsymbol \epsilon (\textbf{k})$ and $\exp \left[i\textbf{k} \cdot \textbf{R}_{l}\right]$, both of which contribute to the calculation of the PAM.
	In the former, the phase difference of $\epsilon (\textbf{k})$ leads to the spin PAM, which is quantised as an integer.
	The latter factor, $\exp \left[i\textbf{k} \cdot \textbf{R}_{l}\right]$
	provides the orbital PAM, which is therefore equal to $\dfrac{1}{2\pi}\textbf{k} \cdot \textbf{c}$ under a three-fold helical rotation.
	Consequently, for each wavenumber $\textbf{k}$, the total PAM is
	\begin{equation}
	m_{\text{PAM}}=m_{\text{PAM}}^{\text{s}}+m_{\text{PAM}}^{\text{o}}.  \label{totalPAM}
	\end{equation}
	Here, we considered only the spin PAM in the phonon dispersion, as shown in Fig. 5(a).

	\noindent \textbf{Calculation of AM.}
	The AM of phonons at wavenumber $\textbf{k}$ of mode number $j$ is defined as\cite{Zhang14}
	\begin{equation}
	m_{\text{AM}}(\textbf{k}, j)=\left(\boldsymbol \epsilon (\textbf{k}, j) ^{\dagger }M\boldsymbol \epsilon (\textbf{k}, j) \right) \hbar.
	\label{LAM}
	\end{equation}
	where $M=\left(
	\begin{array}{ccc}
	0 & -i & 0 \\
	i & 0 & 0 \\%
	0 & 0 & 0
	\end{array}%
	\right) \otimes I_{n\times n} $, where the basis of the $3\times3$ matrix is represented using the orthogonal bases $(u_x, u_y, u_z)$, $n$ is the number of atoms in a unit cell, and $I_{n\times n}$ is a unit matrix of $n\times n$.
	The eigenvector of the dynamical matrix $\boldsymbol \epsilon (\textbf{k}, j)$ is normalised as:
	\begin{equation}
	\boldsymbol \epsilon (\textbf{k}, j)^\dagger \boldsymbol \epsilon (\textbf{k}, j) = 1.
	\end{equation}

	\noindent
	\textbf{The origin of spin and orbital PAMs. } Here we provide an
	explicit form of the spin and orbital PAMs associated with the screw
	operation $\left\{ C_{3}~|~\textbf{c}/3\right\} $, where $\textbf{c}/3$ is the non-primitive translation. The atomic displacement vector field $\textbf{u}_{\textbf{k}}(\textbf{r})$ associated with the phonon propagation may be generally written as
	\begin{equation}
	\textbf{u}_{\textbf{k}}(\textbf{r})=\boldsymbol \epsilon (\textbf{k}
	)\exp \left( i\textbf{k}\cdot \textbf{r}\right) .
	\end{equation}
	Then, $\textbf{u}_{\textbf{k}}(\textbf{r})$ is
	transformed in accordance with a general rule:
	\begin{equation}
	\left\{ C_{3}~|~\textbf{c}/3 \right\} \textbf{u}_{\textbf{k}}(\textbf{r})=\left[ C_{3}\boldsymbol{\epsilon}(\textbf{k})\right] \exp \left( i\textbf{k}\cdot \left\{ C_{3}~|~ \textbf{c}/3 \right\} ^{-1}\textbf{r}\right) .
	\end{equation}
	Note that $\left\{ C_{3}~|~\textbf{c}/3 \right\} ^{-1}=\left\{C_{3}^{-1}~|~-C_{3}^{-1}\textbf{c}/3\right\} $, $\textbf{k}\cdot \textbf{r}$ is a scalar, and $C_{3}\boldsymbol{\epsilon}(\textbf{k})=\exp \left( -i\frac{2\pi }{3}m\right) \boldsymbol{\epsilon }(\textbf{k})$, thus we have
	\begin{equation}
	\left\{ C_{3}~|~\textbf{c}/3\right\} \textbf{u}_{\textbf{k}}(\textbf{r})=\exp \left[ -i\frac{2\pi }{3}\left(m_{\text{PAM}}^{\text{s}}+m_{\text{PAM}}^{\text{o}}\right) \right] \textbf{u}_{\textbf{k}}(\textbf{r})
	\end{equation}
	where the respective spin and orbital PAMs are introduced by
	\begin{equation}
	m_{\text{PAM}}^{\text{s}}=  m,
	\end{equation}
	and
	\begin{equation}
	m_{\text{PAM}}^{\text{o}}= - \frac{3}{2\pi }\left\{ \left( C_{3}\textbf{k}\right) -\textbf{k}\right\} \cdot \textbf{r} + \frac{1}{2\pi }\left( C_{3}\textbf{k}\right) \cdot \textbf{c},
	\end{equation}
	which yield the total PAM, $m_{\text{PAM}}=m_{\text{PAM}}^{\text{s}}+m_{\text{PAM}}^{\text{o}}$.

	In this study, we consider path $\Delta $ from $\Gamma $ to A,
	where $C_{3}\textbf{k}=\textbf{k}$ denotes the modulo reciprocal lattice vectors. Thus, 
	$m_{\text{PAM}}^{\text{o}}= \frac{1}{2\pi} \textbf{k} \cdot \textbf{c}$.
	The same symmetry properties apply to the photons propagating along the helical axis.
	Then, we obtain the selection rule (\ref{Umklapp}), which includes only 
	$m_{\text{PAM}}^{\text{s}}$,
	by cancelling out the $m_{\text{PAM}}^{\text{o}}$ terms because of the law of conservation of momentum.
	
	\subsection{Data availability}
	Source data are provided with this paper.
    
    \subsection{Code availability}
    All custom codes used for the data processing and numerical simulations are available from the corresponding author upon reasonable request.

\end{document}